# Improving disruptive research in the EU: why strengthening European Research Council grants alone is not enough

Alonso Rodríguez-Navarro

*Departamento de Biotecnología-Biología Vegetal, Universidad Politécnica de Madrid, Avenida Puerta de Hierro 2, 28040, Madrid, Spain*

E-mail address: alonso.rodriguez@upm.es ORCID 0000-0003-0462-223X

## Abstract

Disruptive innovation in the EU is not sufficiently competitive; this weakness puts at risk the social benefits that its citizens take for granted. This report argues that, in addition to addressing structural and economic deficiencies, the EU must improve disruptive research to strengthen its disruptive innovation capacity. Currently, the level of disruptive research is too low. Using graphene research as an example, for which the EU has a specific programme, this report shows that Germany, France, Italy, and Spain cannot compete with Singapore. Even more concerning, the research funded by the European Research Council on graphene fails to compete with research conducted in Singapore. Similarly, the EU is far from competing with the USA or China. A few examples in this report and cited references evidence that the situation is similar in other technologies. To overcome this situation, the EU must adopt drastic changes in research policy. However, such changes face a vanity culture among policymakers and, perhaps, scientists who have been proclaiming an inexistent research excellence for decades. Without drastic changes, the prospect of the EU becoming a technological leader at the level of the USA and China cannot be considered realistic.

## Introduction

"If the European Union keeps stagnating, our countries will not be able to maintain the things its citizens take for granted: unemployment benefits, free healthcare, lifelong pensions, and affordable education" (Garicano et al., p. 3). Among the necessary measures to break the stagnation, there is agreement on the need to escape the *middle technology trap* (Fuest et al., 2024). Otherwise, the prospects for escaping the trap are dubious, which would mean adversity because currently "no frame that implies a promise of the EU becoming a worldwide technological leader can be considered realistic" (Szczygielski et al., 2025, p. 15).

To escape from the *middle technology trap*, aside from the necessary measures to strengthen country integration and investment (Draghi, 2024; Letta, 2024), disruptive research[1] needs to be improved, because EU research is currently too heavily focused on 'normal' research (Rodríguez-Navarro, 2026)—this notion is similar to that proposed by Thomas Kuhn (1970) for 'normal' and 'revolutionary' research. In fact, the importance of innovations cannot be greater than that of the research outputs from which they stem, since the importance of science and patents is strongly associated (Poege et al., 2019).

Despite the crucial importance of improving disruptive research, the European Commission has never launched a project to improve this type of research. The only proposal in this direction is to "double the support for fundamental research through the ERC" (European Commission, 2024a), as proposed in the Draghi Report (Draghi, 2024, p. 29). However, although this proposal is reasonable and consistent with the notion of excellence associated with the European Research Council (ERC)-funded research, support for this excellence comes exclusively from the ERC, without empirical data supporting this notion. For example, statements such as "the ERC stands as a beacon of excellence, pushing the boundaries of knowledge and fostering breakthroughs that not only advance science but also shape the future of society" and "the report highlights that 20% of ERC-funded projects led to breakthroughs, while approximately 58% resulted in major scientific advances" (European Commission, 2024b, pp. 5 and 32) lack empirical support.

Although the low capacity of EU research to compete at the disruptive level is well established (Rodríguez-Navarro, 2025a), little is known about the responsibility of its countries for this deficiency (Rodríguez-Navarro, 2026), and nothing is known about whether this shortcoming can be corrected by increasing funding for the ERC programme. This report seeks to shed light on this uncertainty using approaches that are accessible to readers without specific training in bibliometrics. More technical information can be found in the cited references.

### *Disruptive research and its measurement*

Assuming, as already explained, that disruptive innovations depend on disruptive research, the challenge is to determine how to measure disruptive research, which constitutes an insignificant proportion of 'normal' research. From a statistical point of view, disruptive papers are very highly cited papers that occur at a frequency of approximately 0.01% (Bornmann et al., 2018; Poege et al., 2019). Luckily, research systems are complex systems that can be mathematically characterized, allowing the frequency of very highly cited papers to be calculated from the distribution of low- or medium-cited papers (Rodríguez-Navarro & Brito, 2019).

[1] The terms *disruptive*, *radical*, *breakthrough*, and *discontinuous*, when applied to innovations, are equivalent and refer to innovations that "are associated with new technologies that cause a shift in the technological paradigm and business routines; they create new products that eventually lead to demise of existing products" (Feder, 2018, p. 186). Similar considerations apply to research.

## Methods

This report is based on citation counts, which were obtained using the methods already described (Rodríguez-Navarro, 2025a), with publication and citation windows specific to this report. The frequent delayed recognition of disruptive papers (Wang et al., 2017), the low number of ERC papers, and the interest in presenting a picture that is not excessively distant from the present make it impossible to define flawless publication and citation windows. The selection of the 2016–2020 publication window and the 2023–2025 citation window is not perfect, but these windows proved to be the most convenient for the purpose of this report. They are used in all cases except when showing the progress of China. In that case, the publication and citation windows are indicated in the corresponding figure.

Another characteristic of this report is that all data refer to domestic papers—that is, papers in which all authors belong to the same country and, in the case of the EU, considering the EU as a country. This important matter will be considered below before the conclusions of this report.

To obtain information about the structure of research in countries and their potential to produce disruptive research, I selected four technologies: graphene, solar cells/photovoltaics, lithium batteries, and fuel cells. Graphene was selected because of its importance in many technological fields—the EU has a specific programme, the Graphene Flagship, focused on this research topic. The other three topics were selected because of their importance for addressing climate change, a field of crucial importance. Research on graphene is also used to illustrate the methods applied to measure disruptive research.

## The ERC and NSF programmes in the EU and the USA

Table 1 depicts the number of papers published by the EU and the USA and the proportion of these papers reporting funding from the ERC and NSF programmes, respectively. The first and most obvious conclusion that can be drawn from the data is that the ERC programme in the EU is proportionally much smaller than the NSF programme in the USA. The ratio of the proportions of publications in the USA and the EU funded by the National Science Foundation (NSF) and the ERC, respectively, is approximately ten to one, except for the total number of papers and graphene, where it is approximately seven to one.

Table 1. Importance of ERC and NSF funding of research in the EU and the USA considering the number of papers published

| | Total papers | Graphene | Solar cells | Lithium batteries | Fuel cells |
|---|---|---|---|---|---|
| **EU** | 1,285,123 | 9,049 | 10,978 | 4,492 | 5,576 |
| **EU-ERC** | 30,853 | 628 | 440 | 147 | 128 |
| **USA** | 1,000,617 | 5,802 | 5,798 | 4,426 | 3,477 |
| **USA-NSF** | 168,717 | 2,791 | 2,666 | 1,527 | 1,034 |
| **ERC/all EU (%)** | 2.4 | 6.9 | 4.0 | 3.3 | 2.3 |
| **NSF/all USA (%)** | 16.9 | 48.1 | 46.0 | 34.5 | 29.7 |

## Graphene: a case study

*Parametric approaches*

Table 2. Number of publications on graphene

| World | EU | USA | Germany | France | Italy | Spain | Singapore | EU-ERC | USA-NSF |
|---|---|---|---|---|---|---|---|---|---|
| 146,025 | 9,049 | 5,802 | 1,036 | 669 | 1,023 | 1,119 | 618 | 628 | 2,791 |

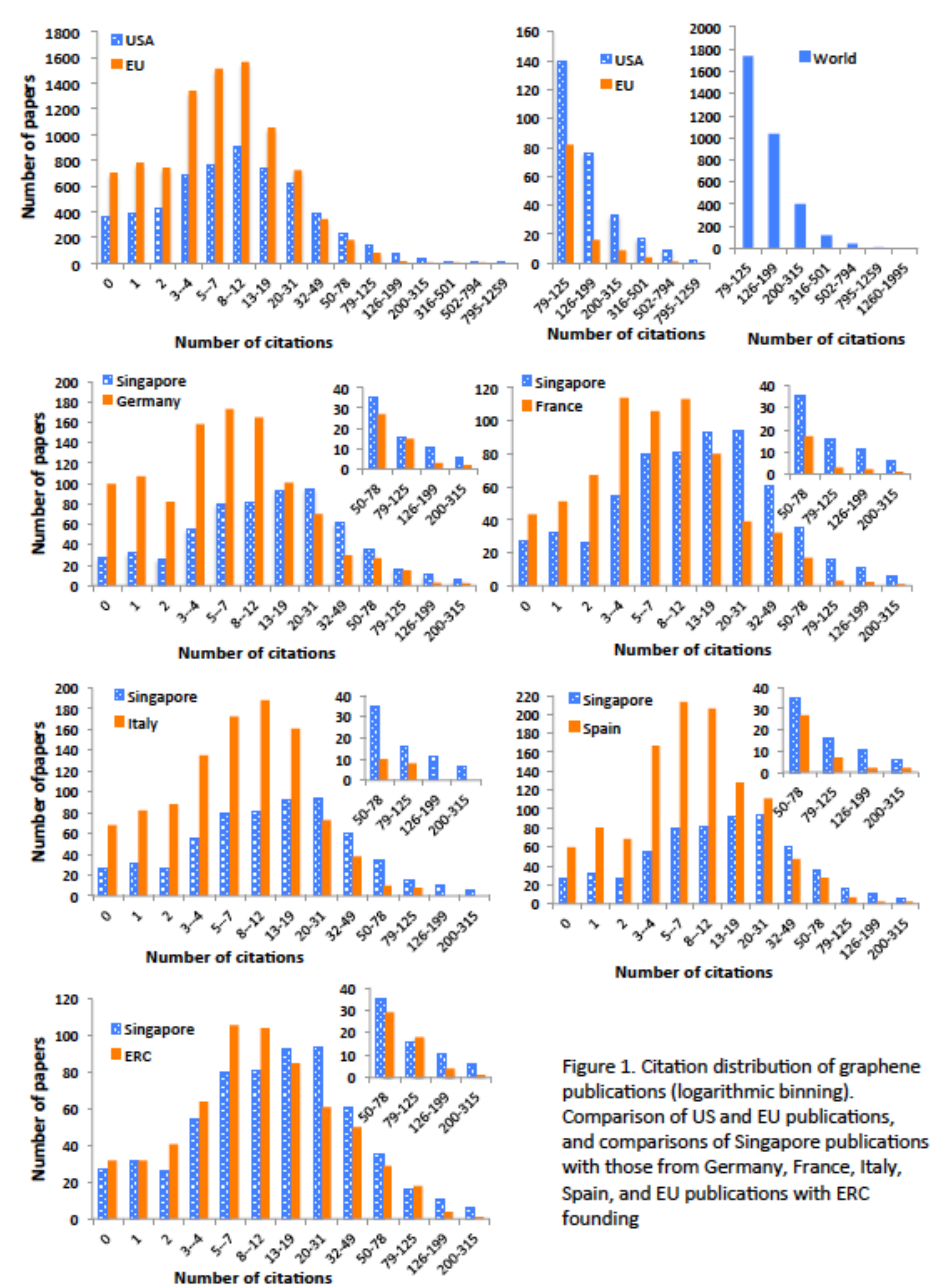

Figure 1. Citation distribution of graphene publications (logarithmic binning). Comparison of US and EU publications, and comparisons of Singapore publications with those from Germany, France, Italy, Spain, and EU publications with ERC founding

Parametric approaches to research assessment are based on citation distributions, which follow a lognormal distribution plus another distribution for uncited and lowly cited papers (Rodríguez-Navarro, 2025b). Lognormal distributions have a long upper tail that cannot be easily shown, but the logarithms of the data, or plots of the data with logarithmic binning, produce approximately normal distributions whose characteristics can be visually perceived. Table 2 depicts the total number of papers in the studied cases, and Figure 1 shows the whole distributions and details of the upper tails.

By comparing the plots of the citation distributions of graphene publications from the USA and the EU, it can easily be observed that the upper tail of the USA plot extends further toward highly cited papers than that of the EU plot. Although the EU publishes many more papers than the USA (9,049 versus 5,802), the 502–794 citation bin contains nine papers from the USA and only one from the EU. EU publications dominate among low- and moderately cited publications, up to 30 citations, but are underrepresented among highly cited publications. This is the 'normal' research trap, which cannot be detected without studying the upper tail of the distribution.

To make similar comparisons with individual EU countries—Germany, France, Italy, and Spain—and with EU-ERC papers, I used Singapore publications as a benchmark because Singaporean research is efficient, possibly almost at the level of the USA. Furthermore, the numbers of publications from Singapore and the other countries are not very different (Table 2), which is necessary for an easier comparison.

Considering the citation levels of the most cited papers from the USA, it can be concluded that research from Singapore, individual countries, and EU-ERC-funded research does not reach the disruptive level. Compared with Singapore, publications from the individual countries follow the same pattern as the EU publications compared with those of the USA: a dominance in low and moderately cited papers[2], while the citation distributions of the EU-ERC-funded and Singapore research are more similar. However, considering the upper tails of the distributions, the papers from Singapore are the most highly cited. Therefore, giving the excellence associated with ERC-funded research in the EU, the results suggest that the best science in the EU lags behind that of Singapore.

*Nonparametric approaches*

Very skewed distributions, such as citation distributions, are more accurately analysed using nonparametric than parametric approaches. For this purpose, publications are ordered from the most to the least cited, and their ranks are subsequently used instead of the number of citations. With this approach, each publication has two ranks: a global rank and a country rank; papers potentially disruptive have very low global ranks. Table 3 shows the 10 most cited publications from the EU and the USA, as well as the last publication in the global top 1% of most cited publications—1,460 publications.

[2] The scientific significance of uncited and lowly cited papers is complex (Rodríguez-Navarro, 2025b) and its treatment is beyond the scope of this report.

Table 3. Global and local ranks of the 10 most cited papers in the USA and EU, and NSF- and ERC-funded papers on graphene. The lower shaded line indicates the ranks of the last paper in the top 1% most cited papers.

| USA | | USA-NSF | | EU | | EU-ERC | | Singapore | | Germany | |
|---|---|---|---|---|---|---|---|---|---|---|---|
| Global | Local | Global | Local | Global | Local | Global | Local | Global | Local | Global | Local |
| 3 | *1* | 3 | *1* | 33 | *1* | 33 | *1* | 14 | *1* | 33 | *1* |
| 5 | *2* | 5 | *2* | 75 | *2* | 317 | *2* | 257 | *2* | 124 | *2* |
| 11 | *3* | 11 | *3* | 81 | *3* | 1273 | *3* | 343 | *3* | 159 | *3* |
| 13 | *4* | 27 | *4* | 124 | *4* | 1311 | *4* | 410 | *4* | 387 | *4* |
| 26 | *5* | 35 | *5* | 159 | *5* | 1,448 | *5* | 432 | *5* | 522 | *5* |
| 27 | *6* | 39 | *6* | 196 | *6* | | | 440 | *6* | 605 | *6* |
| 35 | *7* | 51 | *7* | 265 | *7* | | | 563 | *7* | 1182 | *7* |
| 39 | *8* | 54 | *8* | 317 | *8* | | | 632 | *8* | 1,311 | *8* |
| 44 | *9* | 57 | *9* | 320 | *9* | | | 722 | *9* | | |
| 49 | *10* | 69 | *10* | 364 | *10* | | | 744 | *10* | | |
| | | | | | | | | | | | |
| 1,459 | *125* | 1,453 | *60* | 1,448 | *27* | | | 1,442 | *17* | | |

Another advantage of the double-rank analysis is that it allows the determination of a size-independent parameter that measures research efficiency. This is possible because the double ranks of the most cited papers follow a power law. Figure 2 shows the double-rank plots for the top 1% of most cited papers[3], for which the exponent can be calculated using several methods (Rodríguez-Navarro & Brito, 2019).

Table 4 summarizes the EU-USA comparison. The exponent of the power law, which measures research efficiency—the lower the exponent, the higher the efficiency—has been calculated for the papers included in the top 1% most cited papers.

Despite having far fewer publications (Table 1), the USA-NSF-funded research is similarly disruptive to the total US research (Table 3 and Figure 2)—US papers with the lowest global ranks are also in the USA-NSF list.

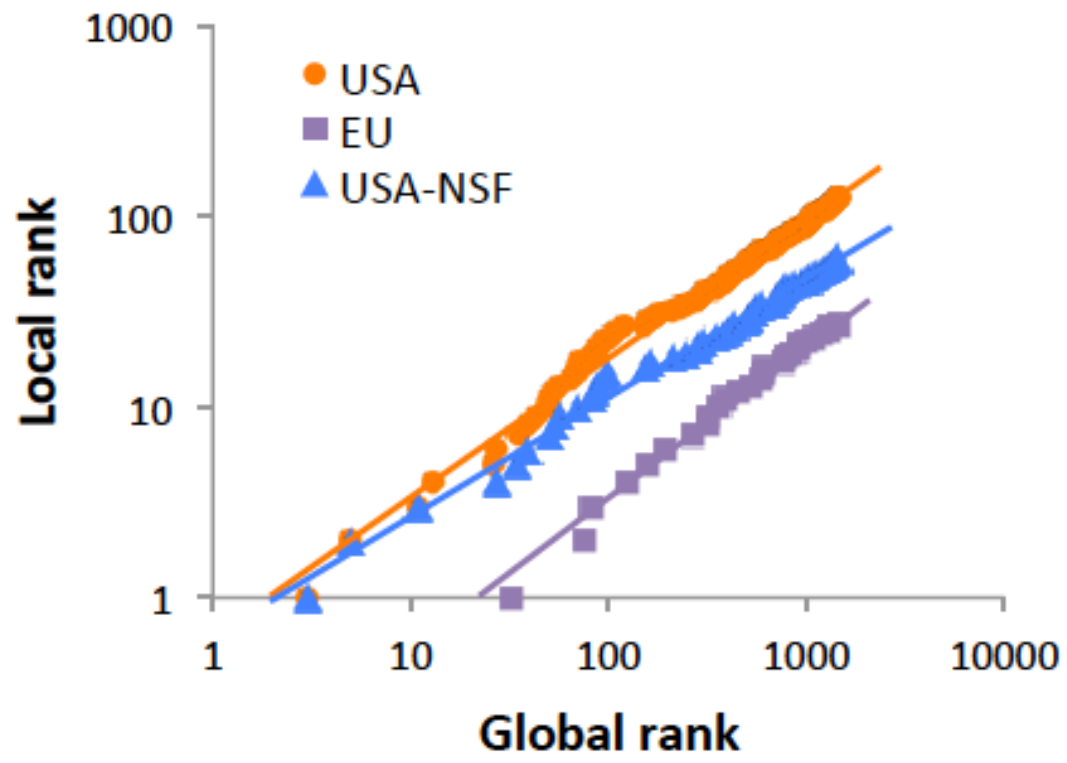

Figure 2. Double rank plots of publications on graphene from the USA, the EU, and the USA with NSF funding

[3] Log-log plots of power-law-distributed data are straight lines.

Table 4. Characterization of highly cited research on graphene. Total number of papers, number of top 1% most cited papers, and exponent of the double rank power law corresponding to the top 1% most cited papers (Figure 2)

| | Total | Top 1% | Exponent |
|---|---|---|---|
| **USA** | 5802 | 125 | 0.72 |
| **USA-NSF** | 2791 | 60 | 0.60 |
| **EU** | 9049 | 27 | 0.85 |
| **EU-ERC** | 628 | 5 | n.c. |

n.c., the low number of EU-ERC papers does not allow the power law to be fitted

In summary, the citation study of EU and US publications on graphene demonstrates that EU research on graphene is far from competing with US research and that, at the disruptive level, EU-ERC publications make no contribution.

*The rise of China*

Table 5. Evolution of the number of publications on graphene

| | 2015–2016 | 2018–2019 | 2021–2022 |
|---|---|---|---|
| **China** | 16,278 | 27,614 | 29,267 |
| **EU** | 3,097 | 3,657 | 3,793 |
| **USA** | 2,392 | 2,302 | 1,859 |

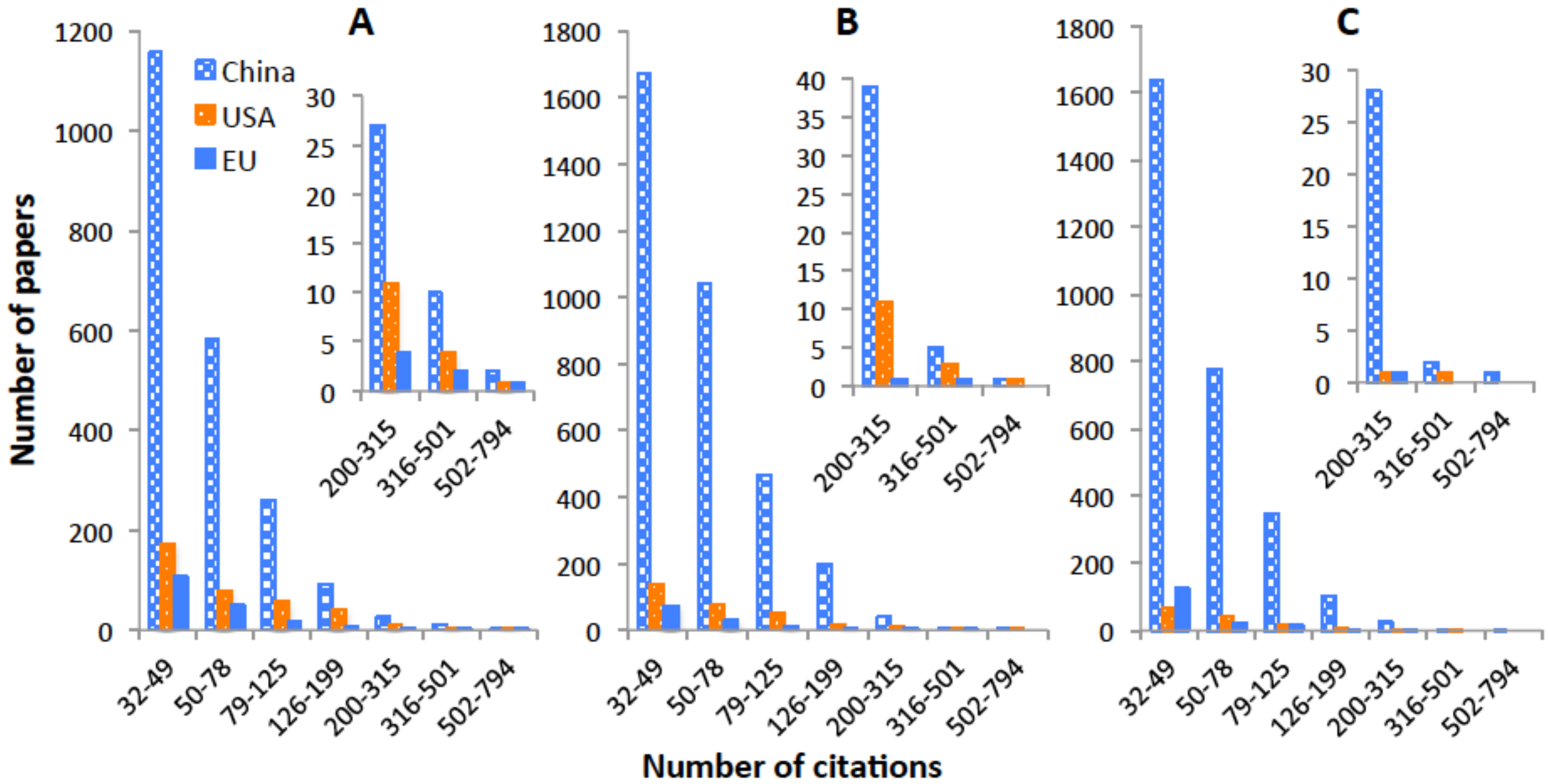

Figure 3. Upper tails of citation distributions of graphene publications from China, the USA, and the EU (logarithmic binning) Publication and citation windows: A, 2015–2016 and 2018–2019; B, 2018–2019 and 2021–2022; C, 2021–2022 and 2024–2025.

The aim of this report is to assess the efficiency of EU research in comparison with that of the USA. This comparison is straightforward because the numbers of publications are not very different, which avoids the need for any kind of calculation. This is not the case for China. However, excluding China would prevent us from understanding how global research is currently evolving.

To address this problem, Table 5 depicts the evolution of the number of publications on graphene, and Figure 3 shows the upper tails of citation distributions in three periods: 2015–2016, 2018–2019, and 2021–2022. The results suggest a strong dominance of China. However, because the citation windows that can be used for these periods are not the most convenient for assessing disruptive research (Wang et al., 2017), the apparent dominance of China over the USA must be interpreted with caution. In contrast, considering the previous result demonstrating the inability of the EU to compete with the USA, it seems clear that, after the 2018–2019 period, the EU is far from being able to compete with China.

## Solar cells, lithium batteries, and fuel cells

The citation distributions of publications on solar cells, lithium batteries, and fuel cells (Figure 4) provide a picture consistent with the study of graphene publications. Except in lithium batteries, the EU and US citation distributions in the other two technologies are different because a larger number of publications in the EU than in the USA accumulate in low and medium number of citations. As in graphene, the EU and US upper tails of citation distributions are notably different, and the difference between the numbers of top 1% most cited papers is large in all cases (Table 6). In lithium batteries the total numbers of papers from the EU and the USA are similar (Table 1) and, except in the upper tail, the citation distributions are also similar.

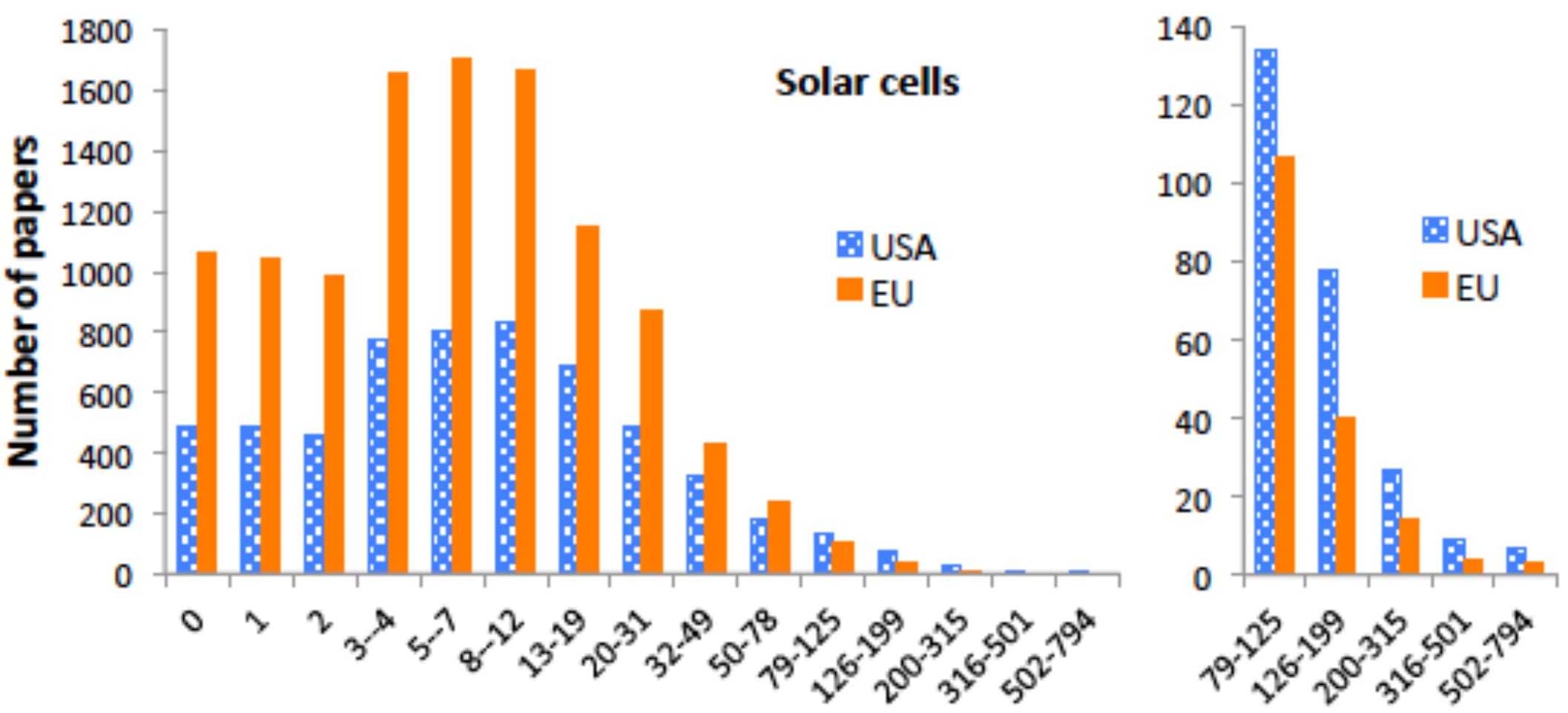

Figure 4. The caption for the figure is on the next page

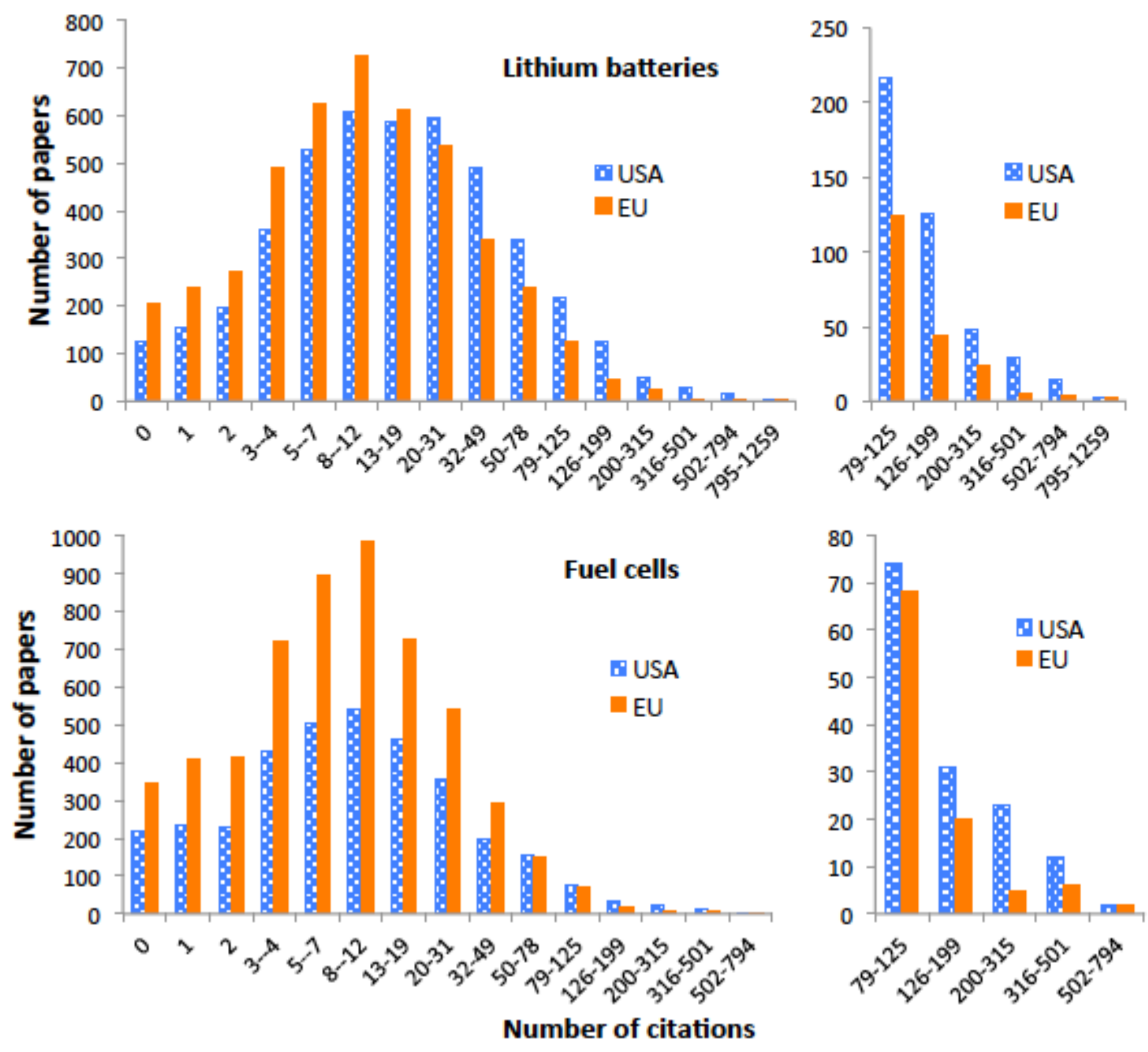

Figure 4. Citation distribution of publications on solar cells, lithium batteries, and fuel cells (logarithmic binning). Comparison of US and EU publications

Table 6. Global and local ranks of the 10 most cited papers from the USA and EU, and NSF- and ERC-funded papers in the indicated technological topics. The lower shaded lines indicate the ranks of the last paper in the top 1% most cited papers

| Solar cells | | | | | | | |
|---|---|---|---|---|---|---|---|
| **USA** | | **USA-NSF** | | **EU** | | **EU-ERC** | |
| **Global** | **Local** | **Global** | **Local** | **Global** | **Local** | **Global** | **Local** |
| 2 | *1* | 33 | *1* | 18 | *1* | 141 | *1* |
| 24 | *2* | 34 | *2* | 23 | *2* | 247 | *2* |
| 26 | *3* | 36 | *3* | 27 | *3* | 428 | *3* |
| 33 | *4* | 45 | *4* | 48 | *4* | 543 | *4* |
| 34 | *5* | 56 | *5* | 68 | *5* | 679 | *5* |
| 36 | *6* | 61 | *6* | 72 | *6* | 780 | *6* |
| 44 | *7* | 70 | *7* | 79 | *7* | | |
| 45 | *8* | 88 | *8* | 141 | *8* | | |
| 56 | *9* | 90 | *9* | 154 | *9* | | |
| 59 | *10* | 110 | *10* | 180 | *10* | | |
| | | | | | | | |
| 865 | *127* | 861 | *67* | 814 | *50* | | |
| Table 6 continues on the next page | | | | | | | |

| Lithium batteries | | | | | | | |
|---|---|---|---|---|---|---|---|
| USA | | USA-NSF | | EU | | EU-ERC | |
| Global | Local | Global | Local | Global | Local | Global | Local |
| 1 | *1* | 1 | *1* | 5 | *1* | 25 | *1* |
| 2 | *2* | 2 | *2* | 6 | *2* | 207 | *2* |
| 9 | *3* | 15 | *3* | 25 | *3* | 320 | *3* |
| 11 | *4* | 28 | *4* | 34 | *4* | 519 | *4* |
| 12 | *5* | 29 | *5* | 44 | *5* | | |
| 15 | *6* | 43 | *6* | 54 | *6* | | |
| 20 | *7* | 47 | *7* | 70 | *7* | | |
| 21 | *8* | 48 | *8* | 107 | *8* | | |
| 22 | *9* | 59 | *9* | 117 | *9* | | |
| 27 | *10* | 60 | *10* | 129 | *10* | | |
| | | | | | | | |
| 532 | *117* | 527 | *48* | 529 | *41* | | |
| | | | | | | | |

| Fuel cells | | | | | | | |
|---|---|---|---|---|---|---|---|
| USA | | USA-NSF | | EU | | EU-ERC | |
| Global | Local | Global | Local | Global | Local | Global | Local |
| 1 | *1* | 5 | *1* | 6 | *1* | 33 | *1* |
| 5 | *2* | 26 | *2* | 10 | *2* | 177 | *2* |
| 9 | *3* | 31 | *3* | 23 | *3* | | |
| 24 | *4* | 32 | *4* | 29 | *4* | | |
| 26 | *5* | 38 | *5* | 33 | *5* | | |
| 28 | *6* | 54 | *6* | 35 | *6* | | |
| 30 | *7* | 73 | *7* | 36 | *7* | | |
| 31 | *8* | 83 | *8* | 44 | *8* | | |
| 32 | *9* | 84 | *9* | 61 | *9* | | |
| 38 | *10* | 106 | *10* | 80 | *10* | | |
| | | | | | | | |
| 417 | *70* | 417 | *18* | 400 | *33* | | |

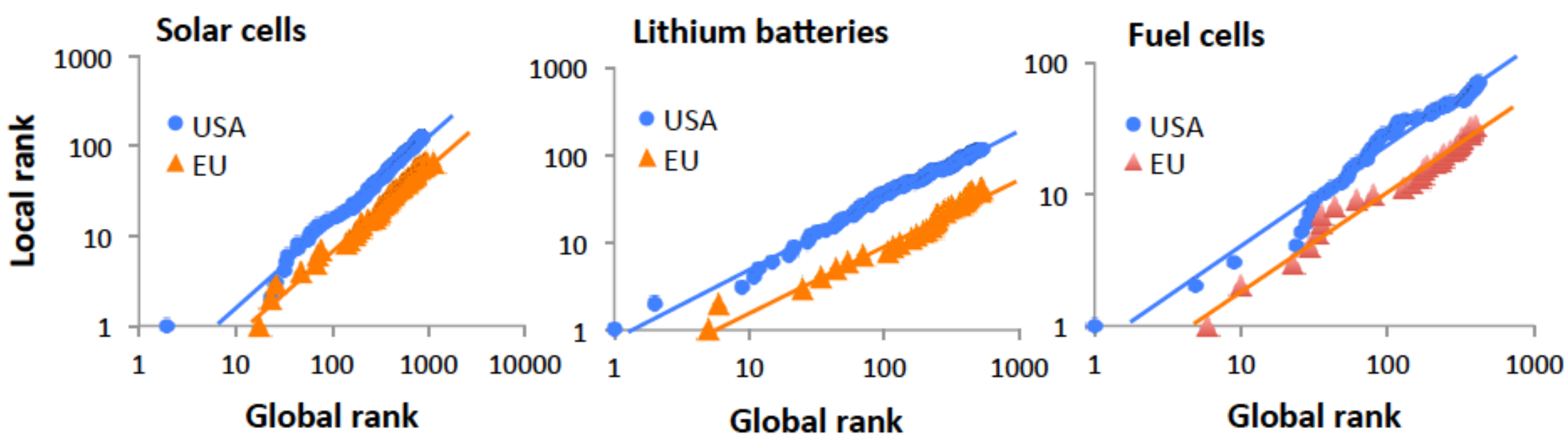

Figure 5. Double rank plots of publications from the USA and the EU in the indicated technological topics

The double-rank plots for the USA and the EU across the three technologies show smaller differences in efficiency than those observed for graphene. The research efficiencies of the top 1% most-cited papers are similar—the exponents of the power laws can be visualized by the slopes of the straight lines in the log-log plots in Figure 5. This observation, together with the lower number of top 1% most-cited papers, indicates that the lower research efficiency of the EU compared with the USA does not apply to the research that produces the top 1% most-cited papers.

As in graphene, in these mature technologies, only a very small proportion of research is competitive, while the inefficiency occurs in the remainder of the research. For example, in solar cells, the numbers of EU and US publications with between zero and 78 citations are 10,806 and 5,540, respectively, whereas the numbers with more than 78 citations are 172 and 258, respectively. Consistent with this, the numbers of top 1% most-cited papers are 50 and 127, respectively (Table 6). All this suggests that the similar efficiency observed in the EU and the USA is reduced to approximately 50 EU publications.

As in graphene, the contribution of ERC-funded papers to pushing the boundaries of knowledge in solar cells, lithium batteries, and fuel cells is irrelevant (Table 6).

## Size and efficiency

Before drawing conclusions from the empirical data described above, an important point to consider is the relationship between size and efficiency. As I have already mentioned, research systems are complex systems that can be mathematically characterized. Highly cited—or disruptive—papers do not occur by chance; they are linked to papers that receive lower numbers of citations. Therefore, each research system has a certain probability of publishing a paper with a number of citations at the disruptive level (Rodríguez-Navarro & Brito, 2019). Serendipity exists, but a country's science policy cannot be based on waiting serendipitous discoveries.

For example, the numbers of papers from Singapore and the ERC-funded programme on graphene are similar (Table 2), but the numbers of papers in the 126–199 and 200–315 citation bins are much higher for Singapore—11 versus 4 and 6 versus 1, respectively. This implies that Singapore is more efficient than the EU-ERC programme, which is consistent with the results shown in Table 3. Despite its high efficiency, Singapore cannot compete with the EU in terms of the overall number of disruptive papers because its research system is too small. If a research system doubles in size, the number of papers corresponding to all the bins shown in Figures 1, 3, and 4 doubles, and the upper tail extends to the right[4].

The same applies to the comparison of China and the USA in graphene research. The US research system is more efficient than that of China (Rodríguez-Navarro, 2026), but the 16-fold difference in their numbers of publications in 2021–2022 (Table 5) explains the differences in the extreme of the upper tail in Figure 3.

A contrasting example is provided by the USA-NSF publications on graphene (Tables 1 and 3, and Figure 2). Despite having half the number of publications, these papers achieve the same level of disruption as US publications. However, the question arises as to whether the successful USA-NSF publications would have been possible without the broader body of US publications.

[4] The mathematical characterization of the relationships between size and efficiency (Rodríguez-Navarro, 2026). is beyond the scope of this report.

From all of the above, it can be concluded that, at the level of disruptive research, a policy aimed at basing the scientific success of a country or the EU on a few highly competitive institutions is unlikely to produce the desired outcome. This is because the probability of making a revolutionary discovery is very small, even in the most competitive institutions. It is worth noting that the winner-takes-all principle applies in disruptive research, implying that competing with China requires not only high efficiency but also a large research system. Neither condition currently applies to the EU.

The most challenging situation for small, highly competitive countries and large, less competitive countries is that they may contribute to the success of the US and China because ‘normal’ science is globalized. And the US and China are the countries that currently best leverage this ‘normal’ science: the former due to its efficiency and the latter due to its size.

None of the above applies to incremental innovations, even those with the greatest market impact. The 'winner-takes-all' principle does not apply to these innovations.

## Domestic and internationally collaborative publications

Another issue to be considered before drawing conclusions in this report is the use of domestic papers to analyse the capacity of the EU to innovate at a disruptive level.

The contribution of the EU to pushing the boundaries of knowledge is made through publications that can be either domestic—i.e., involving only EU countries—or collaborative, involving also external countries. In parametric approaches, the number of citations can be divided among countries according to the number of authors from the EU or from countries external to the EU.

This type of distribution has also been developed in some non-parametric approaches (Perianes-Rodriguez & Ruiz-Castillo, 2015), but not in double rank analysis. To compensate for this limitation, double-rank analysis, as shown in Figures 2 and 5 and Tables 3 and 6, can be applied to domestic and internationally collaborative papers. However, while both types of papers must be considered when studying the contribution to pushing the boundaries of knowledge, estimating the capacity of the EU to generate disruptive innovations is a different problem. In fact, EU collaborative papers that are subsequently transformed into innovations in the USA or China do not represent a success for the EU.

In line with this reasoning, the difference in rank between domestic and collaborative papers in the EU is much greater than in the USA and China. Table 7 shows that, while the EU seems to be a scientific powerhouse considering the total number of papers, this appearance disappears when international collaborations are eliminated, or simply when collaborations with the USA and China are excluded[5]. In domestic papers, the global rank of the 10th-ranked EU paper is 154, quite distant from the ranks associated with potentially disruptive papers. In the case of both the USA and China, eliminating international collaborations has a detrimental effect, but one that is much smaller than in the EU. What Figure 7 shows for solar cells/photovoltaics is similar to what is observed in other technological and biomedical topics (Rodríguez-Navarro, 2025a).

Table 7. Global ranks of the 10 most cited papers from the EU, the USA, and China, considering the total number of publications, and divided into domestic and those with external collaborations. In the case of the EU, a fourth column includes international publications but excluding those in which the collaborating countries are either the US or China.

| **Solar cells** | | | | | | | | | | |
|---|---|---|---|---|---|---|---|---|---|---|
| | **EU** | | | | **USA** | | | **China** | | |
| **Local rank** | **Total** | **Internt Collab.** | **Domest.** | **NOT USA/China** | **Total** | **Internt Collab.** | **Domest.** | **Total** | **Internt Collab.** | **Domest.** |
| ***1*** | 1 | 1 | 18 | 7 | 2 | 5 | 2 | 1 | 1 | 3 |
| ***2*** | 5 | 5 | 23 | 37 | 5 | 9 | 24 | 3 | 5 | 6 |
| ***3*** | 7 | 7 | 27 | 41 | 9 | 11 | 26 | 5 | 8 | 13 |
| ***4*** | 9 | 9 | 48 | 51 | 11 | 15 | 33 | 6 | 9 | 31 |
| ***5*** | 11 | 11 | 68 | 53 | 14 | 14 | 34 | 8 | 14 | 42 |
| ***6*** | 18 | 21 | 72 | 109 | 15 | 16 | 36 | 9 | 15 | 43 |
| ***7*** | 21 | 22 | 79 | 124 | 16 | 17 | 44 | 13 | 16 | 49 |
| ***8*** | 22 | 37 | 141 | 132 | 17 | 21 | 45 | 14 | 17 | 60 |
| ***9*** | 23 | 41 | 154 | 148 | 21 | 28 | 56 | 15 | 22 | 76 |
| ***10*** | 27 | 51 | 180 | 155 | 24 | 32 | 59 | 16 | 28 | 77 |
| | | | | | | | | | | |
| **Graphene** | | | | | | | | | | |
| | **EU** | | | | **USA** | | | **China** | | |
| **Local rank** | **Total** | **Internt Collab.** | **Domest.** | **NOT USA/China** | **Total** | **Internt Collab.** | **Domest.** | **Total** | **Internt Collab.** | **Domest.** |
| ***1*** | 10 | 10 | 33 | 89 | 1 | 1 | 3 | 10 | 2 | 10 |
| ***2*** | 17 | 17 | 75 | 103 | 2 | 2 | 5 | 17 | 12 | 17 |
| ***3*** | 20 | 20 | 81 | 138 | 3 | 4 | 11 | 20 | 15 | 20 |
| ***4*** | 30 | 30 | 124 | 144 | 4 | 6 | 13 | 30 | 18 | 30 |
| ***5*** | 33 | 47 | 159 | 177 | 5 | 8 | 26 | 47 | 24 | 47 |
| ***6*** | 47 | 53 | 196 | 212 | 6 | 10 | 27 | 53 | 25 | 53 |
| ***7*** | 53 | 60 | 265 | 227 | 8 | 12 | 35 | 60 | 28 | 60 |
| ***8*** | 60 | 64 | 317 | 385 | 10 | 14 | 39 | 64 | 31 | 64 |
| ***9*** | 64 | 70 | 320 | 386 | 11 | 15 | 44 | 70 | 32 | 70 |
| ***10*** | 70 | 78 | 364 | 409 | 12 | 17 | 49 | 78 | 36 | 78 |

## Research funding

The last consideration before drawing conclusions concerns the widespread idea that the low proportion of disruptive research in the EU is due to lower R&D investment compared with the USA. Aside from differences in industrial structures, which condition private investments, the existence of this "investment gap" is not clear (Gross et al., 2025). Furthermore, there is no evidence that the concentration of papers in the low and medium citation ranges, as depicted in Figures 1 and 4, is the result of the level of investment. This hypothesis can be ruled out by considering that the same pattern occurs in Germany and Spain, despite substantial differences in R&D investment.

[5] The data for solar cells are representative of many technologies. Graphene is included, but the EU's competitiveness in graphene is very low and does not reflect what is happening in other technologies.

The evidence indicates that the "normal" research trap in the EU and many of its countries is a consequence of misguided research policies (Rodríguez-Navarro, 2026). This includes designing systems that are too large in relation to funding capacity in technological and biomedical fields where research demands high funding.

The EU does not need to invest more to improve its disruptive research. Investing more would increase its scientific size, but increasing efficiency does not require large investments.

## Conclusions

This report deals with the research policy steps that the EU should take to escape the *middle technology trap* (Fuest et al., 2024) and with the question of whether doubling the ERC funding programme would be a successful approach.

The first conclusion is that, to compete with the USA and China in disruptive research, the EU must substantially increase its efficiency. Among the four largest EU countries, Germany has the most efficient research system (Rodríguez-Navarro, 2026), but Germany lags behind Singapore in graphene research (Figure 1). This does not occur in other technologies, but only because the number of papers published in Germany is much higher than in Singapore (Rodríguez-Navarro, 2025a).

Regarding the ERC programme, EU-ERC publications currently make no contribution at the frontier of knowledge. Table 3 shows that the most cited papers on graphene in the USA—globally ranked 3rd, 5th, and 11th—are NSF-funded, and that, among the ten most cited papers in the USA, six are also included in the USA-NSF list. A similar comparison between the EU and EU-ERC lists shows that only the first and second papers in the EU-ERC list are included in the EU list—ranked 33rd and 317th in the global list—which implies that they do not push the frontiers of knowledge.

Doubling the number of ERC publications will not be sufficient to compete with Singapore; without improving the efficiency, it would be necessary to increase the programme fivefold to catch up with it. The ERC programme has two problems: it is too small and its efficiency is too low.

This does not imply that the ERC programme is not potentially important or that strengthening it to the level of the NSF programme in the USA (Table 1) is not a rational objective. However, to be successful, ERC projects should be carried out only in countries with competitive research efficiency, because the success of ERC projects depends on the country in which they are executed (Rodríguez-Navarro & Brito, 2020). If research in the UK and Switzerland were integrated into the EU, most ERC projects should be carried out in these countries and the Netherlands. Most probably, some smaller European countries are also research-efficient, but this should be empirically demonstrated.

A similar misunderstanding applies to the proposal to attract US talent to improve EU research. The EU has enough talent to develop competitive research. Examining the names of the authors of the most cited papers from Singapore on graphene (Figure 1), one can see that Singapore did not need to import US talent to develop competitive research. It is difficult to believe that the EU must resort to importing talent to improve its research. The problem is that EU researchers emigrate.

Regardless of the considerations given to ERC-funded research and to the number of exceptional researchers or institutions, it is worth emphasizing that the underperformance of EU research lies in the bulk of its publications. In the case of solar cells described above, this inefficiency applies to 10,806 of the 10,978 papers. This lack of efficiency in the bulk of research is what limits the EU's capacity for disruptive research and this is the limitation that needs to be eliminated in the EU.

Currently, the most important impediment to improving EU research efficiency is its culture of research vanity, which involves proclaiming nonexistent successes. The proposal of the *European Paradox* in 1995, its reinforcement by a *High Level Group* in 2017, and countless documents proclaiming scientific excellence are products of this culture. In the EU, "illusory successes bewilder research policy" (Rodríguez-Navarro, 2026, p. 28).

To give just two examples of claims that can be shown to be incorrect using the data presented in this report, the *Annual Report 2025 of Graphene Flagship* repeatedly mentions "Europe's global leadership" in graphene, and the *Annual Report on the ERC Activities and Achievements in 2024* is full of references to the scientific excellence of ERC-funded research.

Similarly, regarding the attraction of talent from the USA, the "Choose Europe" programme (https://commission.europa.eu/topics/research-and-innovation/choose-europe_en) uses statements such as "As a world-leading centre for research and innovation with freedom of science, the European Union offers an ideal environment to advance your career" and "Europe is where world-class research becomes real-world impact." These statements are products of the same culture of scientific vanity.

Unless the EU abandons this culture of vanity, which involves proclaiming nonexistent successes, it will not develop a research policy that allows it to compete with the USA and China in disruptive innovation.

There is no doubt that many factors constrain disruptive innovation in the EU and that these constraints must be removed (Draghi, 2024; Letta, 2024). However, if they are removed without improving disruptive research, the goal of improving disruptive innovation will not be achieved. Unless drastic measures are applied to increase the size of the research system in terms of the number of publications and improve its efficiency, the notion that "no frame that implies a promise of the EU becoming a worldwide technological leader can be considered realistic" (Szczygielski et al., 2025, p. 15) will continue to characterize EU innovation at the disruptive level.